# BayesSummaryStatLM: An R package for Bayesian Linear Models for Big Data and Data Science

Alexey Miroshnikov[1], Evgeny Savel'ev[2], Erin M. Conlon[1*]

1 Department of Mathematics and Statistics, University of Massachusetts, Amherst, Massachusetts, United States of America

2 Department of Mathematics, Virginia Polytechnic Institute and State University, Blacksburg, Virginia, United States of America

* E-mail: econlon@mathstat.umass.edu

**Abstract**

Recent developments in data science and big data research have produced an abundance of large data sets that are too big to be analyzed in their entirety, due to limits on either computer memory or storage capacity. Here, we introduce our R package `BayesSummaryStatLM` for Bayesian linear regression models with Markov chain Monte Carlo implementation that overcomes these limitations. Our Bayesian models use only summary statistics of data as input; these summary statistics can be calculated from subsets of big data and combined over subsets. Thus, complete data sets do not need to be read into memory in full, which removes any physical memory limitations of a user. Our package incorporates the R package `ff` and its functions for reading in big data sets in chunks while simultaneously calculating summary statistics. We describe our Bayesian linear regression models, including several choices of prior distributions for unknown model parameters, and illustrate capabilities and features of our R package using both simulated and real data sets.

# Introduction

Due to the developments in data science and the exponential growth of big data in recent years, statisticians are facing new challenges in the analysis of large data sets. Here, big data is defined as data sets that are too large to be analyzed in their entirety, due to limits on either computer memory or storage size. The application areas affected by big data are extensive, and include genomics, sustainability, public health, finance, energy and climatology, among many others. Here, we introduce Markov chain Monte Carlo (MCMC) procedures for Bayesian linear regression models with normally distributed errors that use only summary statistics as input. The summary statistics can be combined over subsets of data, which allows for analyzing large data sets in chunks. Since complete data sets do not need to be read into R at once, this overcomes any physical memory limits of a user. Our methods are implemented through the R [1] software package **BayesSummaryStatLM** (available from the Comprehensive R Archive Network at http://CRAN.R-project.org/), which integrates functions from the R package **ff** [2]. The **ff** functions keep only a portion of a data file in resident memory in R, while simultaneously calculating summary statistics on each portion. The summary statistics are then summed over portions to create full data summary statistics.

Similar methods that use summary statistics for Bayesian linear regression models were developed by Ordonez et al. [3] in a big data setting. In this work, the authors introduce a database management system (DBMS) procedure for obtaining summary statistics for full data sets; they then perform Bayesian linear regression on the summary statistics. In addition, Ghosh and Reiter [4] develop methods for Bayesian linear regression based on summary statistics from data subsets, but not in a big data context; instead, the application areas are privacy and secure sharing of information. The methods of Ordonez et al. [3] and Ghosh and Reiter [4] include only one choice of prior distribution for the unknown model parameters, and do not provide software tools to implement their methods. In contrast, we provide an R package for Bayesian linear regression models for big data that includes several choices of prior distributions for the unknown model parameters. In addition, our package can accommodate data that is located in separate files or that is distributed via networks. It can also analyze data that is updated over time; for example, streamed data. We describe each of these user-defined options of our package and provide examples in the following sections.

Our paper is organized as follows. In the Methods section, we introduce the Bayesian linear regression models and summary statistics. In the following section, we describe the main features and functions of our R package; we also illustrate our package using both simulated and real data sets. Our work is summarized in the Discussion section.

## Methods

**Bayesian linear regression model**

Bayesian linear regression models are used extensively across a wide range of scientific subjects including economics, finance, and the biological, behavioral and social sciences, among many others; the purpose of linear regression is to model a response variable $Y$ using a set of predictor variables $X = (X_1, ..., X_k)$. The normal errors linear regression model is the following (see Carlin and Louis [5], Gelman et al. [6]):

$$Y_i = \beta_0 + \beta_1 x_{1i} + \beta_2 x_{2i} + \ldots + \beta_k x_{ki} + \varepsilon_i, \quad k \text{ predictor variables, } i=1,\ldots,n, \qquad (1)$$

$$\varepsilon_i \sim Normal(0, \sigma^2). \qquad (2)$$

The likelihood is given by the following expression:

$$p(Y \mid \beta_0, \beta_1, \beta_2, \ldots, \beta_k, \sigma^2) = \frac{1}{(2\pi\sigma^2)^{n/2}} \exp\left[-\frac{1}{2\sigma^2}(Y-X\beta)^T(Y-X\beta)\right], \qquad (3)$$

where $Y$ is an $n \times 1$ column vector, $X$ is an $n \times (k+1)$ matrix, and $\beta$ is a $(k+1) \times 1$ column vector. Here, the unknown model parameters are the linear regression coefficients $\beta = (\beta_0, \beta_1, \ldots, \beta_k)'$ and the error variance parameter $\sigma^2$. In the Bayesian framework, we assign prior distributions to $\beta$ and $\sigma^2$ and produce the joint posterior distribution as the product of the likelihood and prior distributions; here, we assume prior independence for $\beta$ and $\sigma^2$. The full conditional posterior distributions for the unknown model parameters are proportional to the joint posterior distribution, treating all other parameters as fixed constants. For the choices of prior distributions for $\beta$ and $\sigma^2$ in our R package, the corresponding full conditional posterior distributions depend on the data only through the summary statistics $X^T X$, $X^T Y$ for $\beta$, and $X^T X$, $X^T Y$ and $Y^T Y$ (and $Y^T X = (X^T Y)^T$) for $\sigma^2$; these summary statistics and prior distributions are described in the next sections, with details in Appendix A. The Gibbs sampler is used to sample from the full conditional posterior distributions; we provide details of the Gibbs sampler in Appendix B.

**Summary statistics for full conditional posterior distributions**

As mentioned above, for each choice of prior distribution for $\beta$, the full conditional posterior distribution depends on the data only through two summary statistics $X^T X$ and $X^T Y$ (in Appendix A Formulas (A2),(A4),(A8)). Similarly, the full conditional posterior distributions for $\sigma^2$ in Appendix A Formulas (A14) and (A16) depend on the data only through three summary statistics $X^T X$, $X^T Y$ and $Y^T Y$. These values can be calculated by combining summaries from subsets of data. For our package, we assume the data is partitioned horizontally by the samples $n$ into $M$ nonoverlapping subsets, such that if $X$ is dimension $n \times \psi$, then the partition is by the following:

$$X = \begin{pmatrix} X_1 \\ X_2 \\ \vdots \\ X_m \end{pmatrix}, \qquad (4)$$

where each $X_m, m = 1,...,M$ has $\psi$ columns. The $Y$ vector is also partitioned horizontally, similarly to Formula (4). The full data summary statistics are calculated as follows, for $m = 1,…,M$ chunks:

$$\text{Full data } X^T X = \sum_{m=1}^{M} X_m^T X_m, \qquad (5)$$

$$\text{Full data } X^T Y = \sum_{m=1}^{M} X_m^T Y_m, \qquad (6)$$

$$\text{Full data } Y^T Y = \sum_{m=1}^{M} Y_m^T Y_m. \qquad (7)$$

The Gibbs sampler is used to sample from all full conditional posterior distributions; see Appendix B for details.

**Prior distributions for $\beta$ and $\sigma^2$**

In this section, we list the choices of prior distributions for $\beta$ and $\sigma^2$ that are provided by our R package. Full details on these prior distributions and the resulting full conditional posterior distributions are provided in Appendix A.

*Prior distributions for $\beta$*

1) Uniform prior for $\beta$.

2) Multivariate Normal prior for $\beta$ with known mean vector $\mu$ and known covariance matrix $C$.

3) Multivariate Normal prior for $\beta$ with unknown mean vector $\mu$ and unknown covariance matrix $C$.

*Prior distributions for $\sigma^2$*

1) Inverse Gamma prior for $\sigma^2$ with known shape and scale parameters.

2) Inverse sigma squared prior for $\sigma^2$, i.e. the Jeffreys prior for $\sigma^2$ (see [5]).

**Using Package BayesSummaryStatLM**

**1 Package overview**

The R package **BayesSummaryStatLM** has two primary functions:

`read.regress.data.ff()`, which reads in big data and simultaneously calculates summary statistics, and `bayes.regress()`, which takes as input the summary statistics from `read.regress.data.ff()` and produces the MCMC posterior samples for the unknown parameters in the Bayesian linear regression model. We describe each function in the following sections.

## 2 Function for calculating summary statistics

The function `read.regress.data.ff()` takes as input the data files with the predictor variables *X* and response values *Y*, and returns the summary statistics $X^TX$, $X^TY$, and $Y^TY$. This function incorporates the R package `ff` and its function `read.table.ffdf()`. The `read.table.ffdf()` function overcomes memory limitations by keeping only a portion of a data set in physical memory as it reads it in chunks. The data sets can be split across multiple files, and the size of the chunks can be specified by the user, in options described in the following section. In addition, for cases when data may be updated over time, this function can update the previously computed summary statistics with newly-provided data. This eliminates the need to re-read and store the complete data set. See Section 4.3 below for an example of updating summary statistics.

### *2.1 Description of function arguments*

For the `read.regress.data.ff()` function, the user specifies the names of the data files and the column numbers for the *X* and *Y* values. The arguments used in a call to the function `read.regress.data.ff()` are the following:

```
read.regress.data.ff(filename, predictor.cols,
    response.col, first.rows, next.rows,
    update.summaries)
```

The descriptions of the primary function arguments are shown here; additional arguments can be found in the documentation for the `read.table.ffdf()` function in the `ff` package and for the `read.table()` function in the base R environment:

**`filename`** A list that contains the names of the data files. The function expects the file names to conform to the same requirements as the standard R functions for reading files, such as `read.table()`. The filename can either point to a file in the filesystem, point to a database connection, or be a URL string pointing to a network resource. Default = NULL.

**`predictor.cols`** A vector of column numbers that contain the predictor variables. Default = NA.

**`response.col`** The column number that contains the response variable. Default = NA.

**`first.rows`** The number of rows to read in the first chunk of data. Default = 100,000.

**`next.rows`** The number of rows to read in the remaining chunks of data. Default = 100,000.

**`update.summaries`** The name of the R object containing previously-calculated summary statistics (if applicable), to be updated with new data. Default = NULL.

*2.2  Description of returned values*

The returned value for the **read.regress.data.ff()** function is a list containing the summary statistics named **xtx**, **xty** and **yty** (for $X^TX$, $X^TY$ and $Y^TY$, respectively) and the total number of data values **numsamp.data**. These values are used as input to the function **bayes.regress()**, described in the next section. Here, the design matrix *X* includes a leading column of 1's for the *y*-intercept term of the Bayesian linear regression model. The user can specify a model without a *y*-intercept term through arguments in the **bayes.regress()** function; in this case, all summary statistics are automatically revised appropriately by the function **bayes.regress()** (details in the next section).

**3  Function for MCMC posterior sampling of Bayesian linear regression model parameters**

The **bayes.regress()** function is used to generate the MCMC posterior samples for the unknown Bayesian linear regression model parameters. This function takes as input the summary statistics calculated by the function **read.regress.data.ff()**. The user specifies the choices of prior distributions, hyperprior distributions and fixed parameter values where required; the user also specifies starting values for unknown model parameters. The **bayes.regress()** function incorporates the package **mvnfast** [7] and its function **rmvn()** for sampling from multivariate normal distributions. The **mvnfast** package generally speeds sampling compared to other packages due to the use of C++ code through the **Rcpp** [8,9] and **RcppArmadillo** [10] packages and the use of efficient numerical algorithms. The arguments used in a call to the function **bayes.regress()** are the following:

```
bayes.regress(data.values = list(xtx, xty, yty,
    numsamp.data), beta.prior, sigmasq.prior, Tsamp.out,
    zero.intercept)
```

The descriptions of the available function arguments are given here; the descriptions of the list arguments are given in Tables 1-4.

**data.values** A list that contains input values **xtx, xty, yty, numsamp.data**; one optional argument and further details in Table 1.

**beta.prior** A list that contains the arguments for the prior for $\beta$; details in Tables 2 and 3.

**sigmasq.prior** A list that contains the arguments for the prior for $\sigma^2$; details in Table 4.

**Tsamp.out** The number of MCMC posterior samples to produce. Default = 1,000.

**zero.intercept** An optional logical parameter with default = FALSE. If **zero.intercept** = TRUE is specified, the linear regression model sets the y-intercept term $\beta_0$ to zero (i.e. removes the $\beta_0$ term from the model); by default, the y-intercept term $\beta_0$ is included in the model.

*3.1 Description of returned values*

The returned value for the `bayes.regress()` function is a list containing the MCMC posterior samples of the unknown Bayesian linear regression model parameters; the number of MCMC posterior samples is equal to the argument `Tsamp.out`. The output names are the following for $\beta$, $\sigma^2$, $\mu$, $C^{-1}$, respectively: `beta`, `sigmasq`, `mu`, `Cinv`. Further analysis, including plotting and creating posterior summary statistics, can be carried out using the CODA R package (Plummer et al. [11]); examples are given below.

**4 Example: Simulation data**

*4.1 Description of the simulation procedure*

Here, we demonstrate our R package by simulating data from the linear regression model (1) with ten predictor variables, with data sample size 100,000,000, and file size 10.43 GB. Note that this file cannot be read into memory in R using standard functions such as `read.table()`, due to physical memory limits. The predictor data matrix $X = (X_1,...,X_{10})$ was simulated from a multivariate normal distribution by the following, where each column vector represents a predictor variable:

$$X \sim \text{Normal}(0, \Sigma). \qquad (8)$$

The variance-covariance matrix $\Sigma$ is defined as follows:

$$\Sigma_{hh} = 1, \ h = 1,...,10; \ \Sigma_{hh'} = \rho, \ h \neq h'. \qquad (9)$$

We use $\rho = 0.2$, so that the predictors $X_h$, $h = 1,...,10$, have moderate correlation levels. The model parameters $\beta$ were simulated from a standard normal distribution; these are the parameters that are estimated in our analysis (see Table 5). The dependent values $y_i$, $i = 1,...,100,000,000,$ were then simulated from the following linear regression model:

$$Y_i = \beta_0 + \beta_1 x_{1i} + \beta_2 x_{2i} + ... + \beta_{10} x_{10i} + \varepsilon_i, \tag{10}$$

$$\varepsilon_i \sim \text{Normal}(0, \sigma^2). \tag{11}$$

The data file was named "sim.multreg". Here, we assign $\sigma^2 = 1$.

### 4.2 Illustration of the R functions for simulation data

We read into the function **read.regress.data.ff()** the data file that contains the simulated *X* and *Y* values, named "sim.multreg". At the R prompt, the user enters the following command:

```
> sim.regress.data <- read.regress.data.ff(filename =
'./sim.multireg', predictor.cols = c(1:10), response.col =
11, first.rows = 100000, next.rows = 100000)
```

The returned value **sim.regress.data** is a list containing the summary statistics **xtx, xty, yty,** and the total number of data values **numsamp.data** = 100,000,000. These values are then used as input to the function **bayes.regress()** in the **data.values** argument, described below.

Next, we illustrate the function **bayes.regress()** for linear regression model (1) with *k* = 10. We first specify the priors for $\beta$ and $\sigma^2$ as separate commands in R; these are read into the R function **bayes.regress()**. Here, we use prior 2) for $\beta$ and prior 1) for $\sigma^2$, with the default arguments. At the R prompt, the user enters the following commands:

```
> beta.prior.2 <- list(type="mvnorm.known", mean.mu =
rep(0.0, dim(xtx)[1]), cov.C = diag(1.0, dim(xtx)[1]))

> sigmasq.prior.1 <- list(type = "inverse.gamma",
inverse.gamma.a = 1, inverse.gamma.b = 1, sigmasq.init = 1)

> sim.beta.sigmasq.out <- bayes.regress(data.values =
sim.regress.data, beta.prior = beta.prior.2, sigmasq.prior
= sigmsq.prior.1, Tsamp.out = 11000, zero.intercept =
FALSE)
```

Note that the **bayes.regress()** function will automatically extract the values **xtx, xty, yty** and **numsamp.data** from the output produced by the **read.regress.data.ff()** function. The returned value **sim.beta.sigmasq.out** is a list containing a matrix of MCMC posterior samples for $\beta$ of dimension = (Tsamp.out, *k*+1), and a vector of MCMC posterior samples for $\sigma^2$ of dimension = (Tsamp.out).

Further analysis of the MCMC posterior samples can be carried out using the R package CODA [11]; we illustrate this package in the following commands. Here, the output created above named **sim.beta.sigmasq.out** is first converted to class "mcmc" using the **mcmc** command. The history and density plots for the marginal posterior distribution for $\beta_1$ are generated using the **plot** command; results are shown in Figure 1. The summary statistics for the marginal posterior distribution for $\beta_1$ are produced using the **summary** command, with results shown below. We discard the first 1,000 MCMC posterior samples for burnin. At the R prompt, the user enters the following commands:

```
> plot(mcmc(sim.beta.sigmasq.out$beta[1001:11000,2]))

> summary(mcmc(sim.beta.sigmasq.out$beta[1001:11000,2]))
```

The returned value for the **summary** command in R is the following:

```
1. Empirical mean and standard deviation for each variable,
   plus standard error of the mean:

      Mean             SD          Naive SE      Time-series SE
   -8.638e-01      9.977e-05      9.977e-07        9.977e-07

2. Quantiles for each variable:
      2.5%      25%      50%      75%     97.5%
    -0.8640  -0.8638  -0.8638  -0.8637  -0.8636
```

The returned value for the 95% posterior equal-tail credible interval limits of (-0.8640, -0.8636) includes the simulated value of -0.8638 for $\beta_1$. Similar commands are used for the remaining model parameters. We show the posterior summary statistics for all $\beta$ values, $\beta_0, \beta_1, ..., \beta_{10}$, and $\sigma^2$ in Table 5; we also show the history and density plots for the marginal posterior distribution of $\sigma^2$ in Figure 1.

### *4.3 Saving and updating summary statistics*

Here, we provide an example of saving and updating summary statistics using the argument **update.summaries** in our **read.regress.data.ff()** function. We also present examples of the built-in functions of the R environment for saving and loading objects in the workspace; these features are used when the R session needs to be terminated and restarted at a later time.

We first repeat the command to read in the simulation data file named "sim.multreg" and create summary statistics:

```
> sim.regress.data <- read.regress.data.ff(filename =
'./sim.multireg', predictor.cols = c(1:10), response.col =
11, first.rows = 100000, next.rows = 100000)
```

The R object named "sim.regress.data" can be saved onto a local directory using the built-in command in R named **save.image()**, as follows:

```
> save.image(file="c:\simregressdata.RData")
```

This R object can be loaded in a later R session using the built-in command `load()`, as follows:

```
> load(file="c:\simregressdata.RData")
```

When new data becomes available in a file named "sim2.multreg" (for example), the summary statistics included in the R object "sim.regress.data" are updated using the `update.summaries` argument of `read.regress.data.ff()` function, as follows:

```
> sim.regress.data.2 <- read.regress.data.ff(filename =
'./sim2.multireg', predictor.cols = c(1:10), response.col =
11, skip = 0, first.rows = 100000, next.rows = 100000,
update.summaries = sim.regress.data)
```

## 5 Example: Real data

### 5.1 Description of real data

Here, we use real-world data for all flights on commercial airlines within the United States during the three-month time period May 2014 through July 2014, provided by the U.S. Department of Transportation [12]. The outcome value of interest is the number of minutes each flight was delayed; this is defined as the actual arrival time minus the scheduled arrival time. We used all flights with arrival delays between 1 and 120 minutes; we thus had 613,992 data values for *Y* for the flight delay information. We use two predictors $X_1$ and $X_2$, which are defined as departure delay in minutes and the time of day the flight departed. The $X$ and $Y$ values are included in the data file named "air.multreg".

*5.2 Illustration of the R package for real data*

Here, we read into the function **read.regress.data.ff()** the data file that contains the $X$ and $Y$ values, named "air.multreg". At the R prompt, the user enters the following command:

```
> air.regress.data <- read.regress.data.ff(filename =
'./air.multireg', predictor.cols = c(2:3), response.col = 1,
first.rows = 100000, next.rows = 100000)
```

The returned value **air.regress.data** is a list containing the summary statistics **xtx, xty, yty** and the total number of data values **numsamp.data** = 613,992. These values are used as input to the function **bayes.regress()** in the **data.values** argument. Here, we use the linear regression model (1) for $k = 2$. The priors 1) for $\beta$ and 2) for $\sigma^2$ are assigned, with the default arguments as described above. The user enters the following commands at the R prompt for the function **bayes.regress()**:

```
> beta.prior.1 <- list(type = "flat")
```

```
> sigmasq.prior.2 <- list(type = "sigmasq.inverse",
sigmasq.init = 1)

> airlines.beta.sigmasq.out <- bayes.regress(data.values =
air.regress.data, beta.prior = beta.prior.1, sigmasq.prior
= sigmsq.prior.2, Tsamp.out = 11000, zero.intercept =
FALSE)
```

The returned value **airlines.beta.sigmasq.out** is a list containing a matrix of MCMC posterior samples for $\boldsymbol{\beta}$ of dimension = (Tsamp.out, $k$+1), and a vector of MCMC posterior samples for $\sigma^2$ of dimension = (Tsamp.out). Similar CODA package commands to those shown above were used to produce posterior summary statistics, posterior history and density plots for $\boldsymbol{\beta}$ and $\sigma^2$. We show plots for $\beta_1$ and $\sigma^2$ in Figure 2, and summary statistics for all model parameters in Table 5.

## 6 Software computational time

Here, we summarize our computational times for reading in data sets, calculating summary statistics and producing the MCMC posterior samples, using a computer with operating system Windows 7 64-bit and an Intel Core i7-4600U CPU 2.1 GHz Processor; complete results are provided in Table 6. For this, we created simulation data using the procedure similar to that described in Section 4.1 above, and removing the *y*-intercept term from the models. We used the number of predictor variables equal to 10, 100 and 1000, and the number of data values ranging from $10^4$ to $10^9$. The corresponding file sizes ranged from approximately 100 MB to 100 GB. For the number of predictors 10 and 100, the computational time for reading in data sets and simultaneously calculating summary statistics took from several seconds for approximately 100 MB ($10^5$ data values), to 4.6 hours for approximately 100 GB ($10^9$ data values). For producing the 11,000 MCMC posterior samples (including 1,000 for burnin), the computational times ranged from a few seconds to approximately one minute for different choices of prior distributions for $\boldsymbol{\beta}$ and $\sigma^2$. When increasing the number of predictors to 1,000, the computational time was considerably longer; reading in data sets while calculating summary statistics took from approximately 20 seconds for 100 MB ($10^4$ data values), to approximately 7.5 hours for 100 GB ($10^7$ data values). For generating 11,000 MCMC posterior samples, the computational time took from approximately 30 minutes to almost 8 hours for $\boldsymbol{\beta}$ prior 3) and $\sigma^2$ prior 2). Note that for $\boldsymbol{\beta}$ prior 3), the MCMC posterior samples are generated for three sets of unknown model parameters, including $\boldsymbol{\beta}, \boldsymbol{\mu}$ and the precision matrix $\boldsymbol{C}^{-1}$, which slows calculations considerably in large dimensions.

**Discussion**

We introduced and demonstrated the R package **BayesSummaryStatLM** for Bayesian linear regression models with MCMC implementation that uses only summary statistics as input. Since the summary statistics can be calculated for data subsets and combined, the complete data set does not need to be read into physical memory at once. As a result, our package analyzes big data sets in chunks while keeping only a portion of a data set in resident memory, through the use of the function **read.table.ffdf()** of the **ff** package.

We derived expressions for the full conditional posterior distributions for several choices of prior distributions for the regression coefficients $\beta$ and error variance $\sigma^2$ of the Bayesian linear regression model, and showed that these posterior distributions depend on the data only through the summary statistics $X^TX$, $X^TY$ for $\beta$ and $X^TX$, $X^TY$ and $Y^TY$ for $\sigma^2$. Gibbs sampling is used to generate the posterior MCMC samples for all unknown parameters, which are then plotted and summarized using the CODA R package. We illustrated our package for both simulation and real data sets, and included computational times for carrying out analyses with various choices of prior distributions for unknown model parameters.

Our R package is well-suited to large data sets that have frequent updates, by providing options for saving and updating summary statistics. These options save computational time by removing the need to reanalyze previous large data sets when new data becomes available. There is also flexibility on entering data sources; the user can provide data from several machines or from multiple data providers, and the package automatically combines the summary statistics. Alternatively, summary statistics can be produced by the user outside of the package. Our software tools are extendable to additional choices of prior distributions, and are readily adaptable for parallelization of Bayesian analysis on clusters.

# Appendix A

**Prior distributions and full conditional posterior distributions for $\beta$ and $\sigma^2$**

Here, we provide details on the prior distributions and the resulting full conditional posterior distributions for $\beta$ and $\sigma^2$. For each choice of prior distribution for $\beta$, the full conditional posterior distribution depends on the data only through two summary statistics $X^T X$ and $X^T Y$ (in Formulas (A2),(A4),(A8)). Similarly, the full conditional posterior distributions for $\sigma^2$ in Formulas (A14) and (A16) depend on the data only through three summary statistics $X^T X$, $X^T Y$ and $Y^T Y$ (and $Y^T X = \left(X^T Y\right)^T$).

*Prior distributions and full conditional posterior distributions for $\beta$*

In this section, we introduce three choices of prior distributions for $\beta$ that are included in the R package.

**1) Uniform prior for $\beta$**

Here, the following prior distribution is assigned to $\beta$, where $\beta$ is dimension $(k+1) \times 1$:

$$\beta \sim \text{Uniform}. \tag{A1}$$

The resulting full conditional posterior distribution for $\beta$ is (see references [5,6,13,14]):

$$\beta \mid \sigma^2, X, Y \sim \text{Normal}_{(k+1)}\left(\text{mean}=\left((X^T X)^{-1}(X^T Y)\right), \text{covariance}=\left(\sigma^2 (X^T X)^{-1}\right)\right). \tag{A2}$$

## 2) Multivariate Normal prior for $\beta$ with known mean vector and known covariance matrix

Here, the prior distribution for $\beta$ is specified by the following:

$$\beta \sim \text{Normal}_{(k+1)}\left(\text{mean} = \mu,\ \text{covariance} = C\right), \text{ where:} \tag{A3}$$

$\mu$ is a known $(k+1)\times 1$ vector, specified by the user,
$C$ is a known $(k+1)\times(k+1)$ matrix, specified by the user; must be symmetric and positive definite.

The corresponding full conditional posterior distribution for $\beta$ is (see [5,6,13,14]):

$$\begin{aligned}\beta \mid \sigma^2, X, Y \sim \text{Normal}_{(k+1)}\Big(&\text{mean} = \left(C^{-1} + \sigma^{-2} X^T X\right)^{-1}\left(C^{-1}\mu + \sigma^{-2} X^T Y\right),\\ &\text{covariance} = \left(C^{-1} + \sigma^{-2} X^T X\right)^{-1}\Big).\end{aligned} \tag{A4}$$

## 3) Multivariate Normal prior for $\beta$ with unknown mean vector and unknown covariance matrix

Here, the following prior distribution is specified for $\beta$:

$$\beta \mid \mu, C \sim \text{Normal}_{(k+1)}\left(\text{mean} = \mu,\ \text{covariance} = C\right), \text{ where:} \tag{A5}$$

$\mu$ is an unknown $(k+1)\times 1$ vector,
$C$ is an unknown $(k+1)\times(k+1)$ matrix; must be symmetric and positive definite.

The hyperprior distribution for $\mu$ is:
$$\mu \sim \text{Normal}_{(k+1)}\left(\text{mean} = \eta,\ \text{covariance} = D\right), \text{ where:} \tag{A6}$$

$\eta$ is a known $(k+1)\times 1$ vector, specified by the user,
$D^{-1}$ is a known $(k+1)\times(k+1)$ precision matrix, specified by the user; must be symmetric and positive definite.

The hyperprior distribution for precision matrix $C^{-1}$ is:

$$C^{-1} \sim \text{Wishart}_{(k+1)}(\text{degrees of freedom} = \lambda, \text{ scale matrix} = V), \text{ where:} \quad (A7)$$

$\lambda$ is the known degrees of freedom, specified by the user,

$V^{-1}$ is a known $(k+1)\times(k+1)$ inverse scale matrix, specified by the user; must be symmetric and positive definite.

The following are the full conditional posterior distributions for $\beta$, $\mu$ and $C^{-1}$ (see [5,6,13,14]); note that for the unknown model parameters $\mu$ and $C^{-1}$, the full conditional posterior distributions do not depend on the summary statistics:

$$\beta \mid \sigma^2, \mu, C^{-1}, X, Y \sim \text{Normal}_{(k+1)}\left(\text{mean} = \left(C^{-1} + \sigma^{-2} X^T X\right)^{-1}\left(C^{-1}\mu + \sigma^{-2} X^T Y\right),\right.$$
$$\left. \text{covariance} = \left(C^{-1} + \sigma^{-2} X^T X\right)^{-1}\right), \quad (A8)$$

$$\mu \mid \beta, \sigma^2, C^{-1}, X, Y \sim \text{Normal}_{(k+1)}\left(\text{mean} = \left(D^{-1} + C^{-1}\right)^{-1}\left(C^{-1}\beta + D^{-1}\eta\right),\right.$$
$$\left. \text{covariance} = \left(D^{-1} + C^{-1}\right)^{-1}\right), \quad (A9)$$

$$C^{-1} \mid \beta, \sigma^2, \mu, X, Y \sim \text{Wishart}_{(k+1)}\left(\text{degrees of freedom} = (1+\lambda),\right.$$
$$\left. \text{scale matrix} = \left(V^{-1} + (\beta-\mu)(\beta-\mu)^T\right)^{-1}\right). \quad (A10)$$

*Prior distributions and full conditional posterior distributions for $\sigma^2$*

Next, we introduce two choices of prior distributions for $\sigma^2$ that are included in the R package.

**1) Inverse Gamma prior for $\sigma^2$ with known shape and scale parameters**

Here, the prior distribution for $\sigma^2$ is:

$$\sigma^2 \sim \text{Inverse Gamma}(a, b), \text{ where } a \text{ is known, } b \text{ is known.} \quad (A11)$$

The resulting full conditional posterior distribution for $\sigma^2$ is (see [5,6,13,14]):

$$\sigma^2 \mid \boldsymbol{\beta}, X, Y \sim \text{Inverse Gamma}\left(\frac{n}{2}+a, \left[\frac{1}{2}(Y-X\boldsymbol{\beta})^T(Y-X\boldsymbol{\beta})+\frac{1}{b}\right]^{-1}\right). \tag{A12}$$

Note that the expression $(Y-X\boldsymbol{\beta})^T(Y-X\boldsymbol{\beta})$ can be respecified by the following expression, which is based on summary statistics $X^T X$, $X^T Y$ and $Y^T Y$ (and $Y^T X = (X^T Y)^T$):

$$(Y-X\boldsymbol{\beta})^T(Y-X\boldsymbol{\beta}) = Y^T Y - \boldsymbol{\beta}^T X^T Y - Y^T X\boldsymbol{\beta} + \boldsymbol{\beta}^T X^T X\boldsymbol{\beta}. \tag{A13}$$

Thus, the full conditional posterior distribution for $\sigma^2$ can be restated as:

$$\sigma^2 \mid \boldsymbol{\beta}, X, Y \sim$$

$$\text{Inverse Gamma}\left(\frac{n}{2}+a,\ \left[\frac{1}{2}(Y^T Y - \boldsymbol{\beta}^T X^T Y - Y^T X\boldsymbol{\beta} + \boldsymbol{\beta}^T X^T X\boldsymbol{\beta})+\frac{1}{b}\right]^{-1}\right). \tag{A14}$$

**2) Inverse sigma squared prior for $\sigma^2$**

Here, the following prior distribution is specified for $\sigma^2$, which is the Jeffreys prior for $\sigma^2$ (see [5]):

$$\sigma^2 \sim 1/\sigma^2,\ \sigma^2 > 0. \tag{A15}$$

The corresponding full conditional posterior distribution for $\sigma^2$ is (see [5,6,13,14]):

$$\sigma^2 \mid \boldsymbol{\beta}, X, Y \sim \text{Inverse Gamma}\left(\frac{n}{2}, \left[\frac{1}{2}(Y^T Y - \boldsymbol{\beta}^T X^T Y - Y^T X\boldsymbol{\beta} + \boldsymbol{\beta}^T X^T X\boldsymbol{\beta})\right]^{-1}\right). \tag{A16}$$

## Appendix B

**The Gibbs sampler algorithm**

Once the full conditional posterior distributions are determined, the Gibbs sampler proceeds as follows in the R package [5,6]. First, the user selects random starting values for unknown model parameters 2,…,m, but not the first one; these are labeled as iteration $t = 0$. For the number of MCMC samples $t$, $t = 1,…,T$, repeat the following steps a), b):

a) Update the first unknown model parameter by sampling from its full conditional posterior distribution, conditioned on the iteration ($t$-1) values of the remaining unknown model parameters.

b) Update each of the remaining unknown model parameters 2,…,m by sampling from their full conditional posterior distributions, conditioned on the remaining parameters being fixed; these will be iteration ($t$) values for parameters already updated, and iteration ($t$-1) values for parameters not yet updated.

The R package updates $\boldsymbol{\beta}$ first, so this vector does not need a starting value. The returned values of the R package are the $T$ samples from the full conditional posterior distributions of the unknown model parameters. Note that the $m$-tuple of samples of all parameters obtained at iteration ($t$) converges in distribution to a draw from the true joint posterior distribution (Carlin and Louis [5]).

**Figure Legends**

**Figure 1.** Results for the simulation data for the posterior marginal densities, based on the R package CODA commands; 10,000 MCMC posterior samples after burnin of 1,000 are used. (a) history plot for $\beta_1$; (b) density plot for $\beta_1$; (c) history plot for $\sigma^2$; (d) density plot for $\sigma^2$.

**Figure 2.** Results for the real airlines data for the posterior marginal densities, based on the R package CODA commands; 10,000 MCMC posterior samples after burnin of 1,000 are used. (a) history plot for $\beta_1$; (b) density plot for $\beta_1$; (c) history plot for $\sigma^2$; (d) density plot for $\sigma^2$.

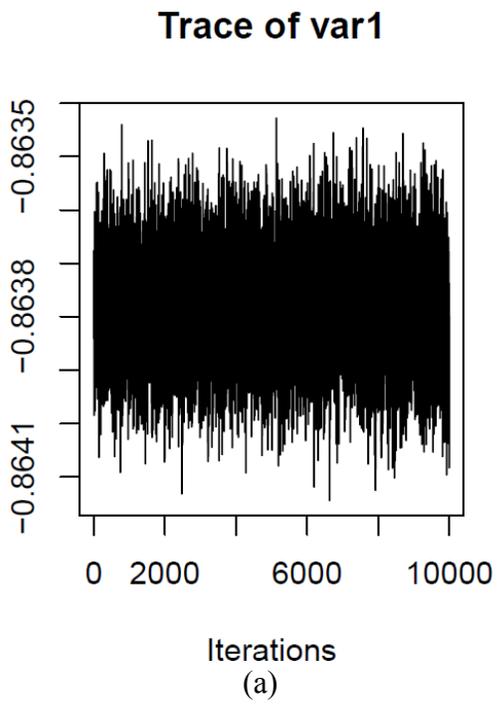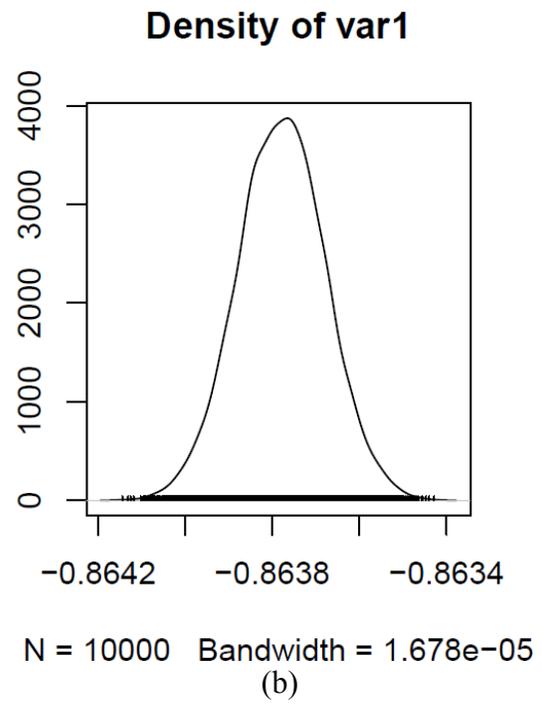

(a) (b)

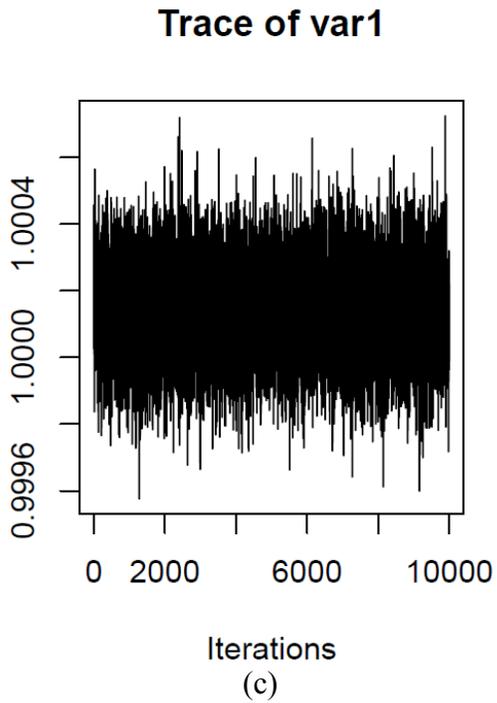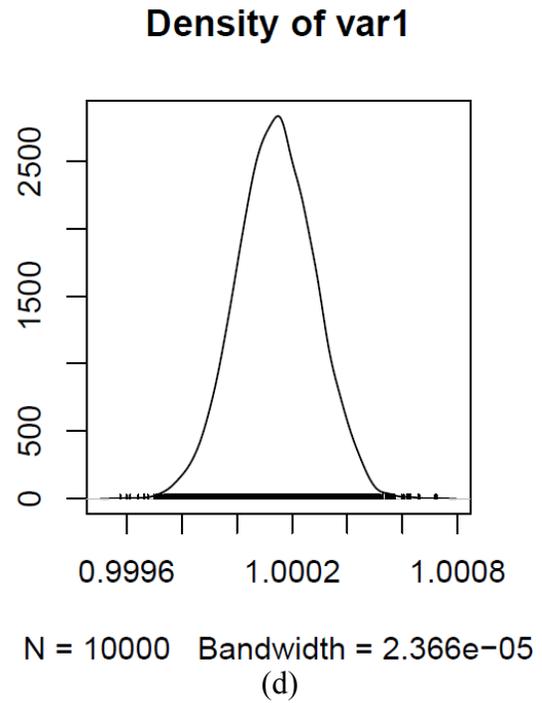

(c) (d)

Figure 1

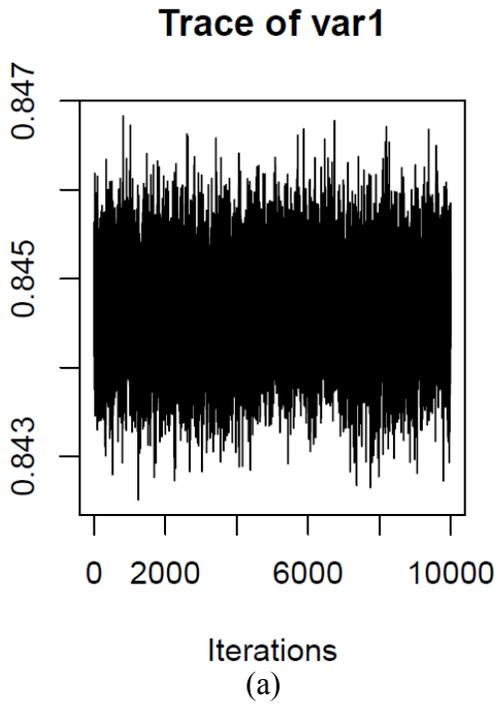
(a)

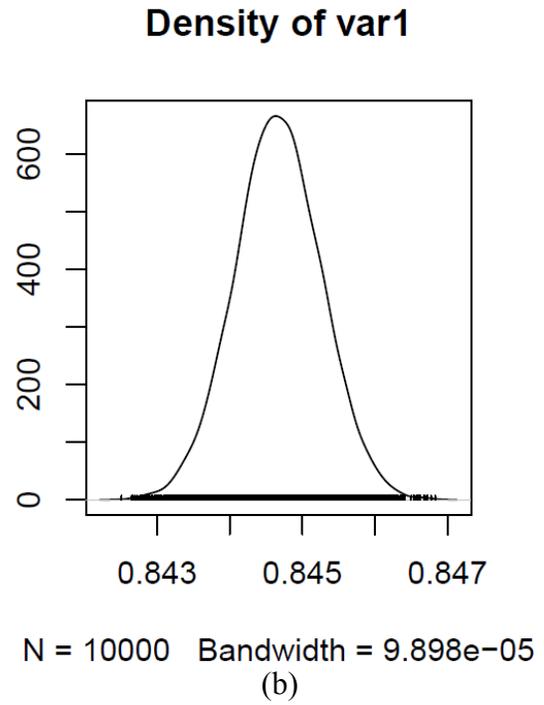
(b)

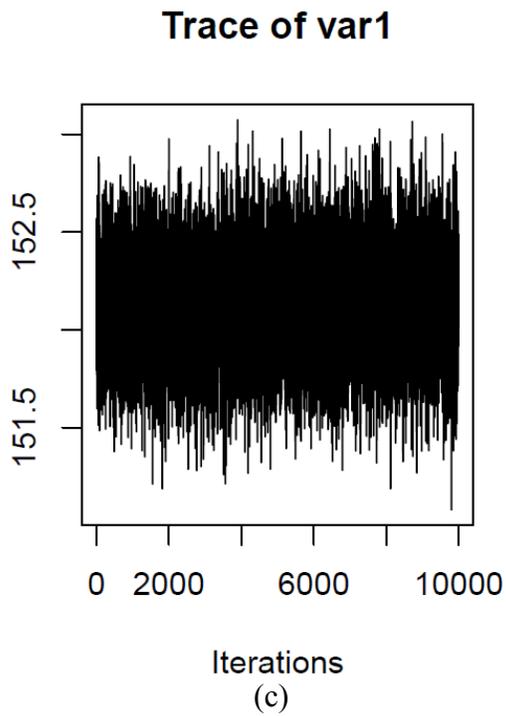
(c)

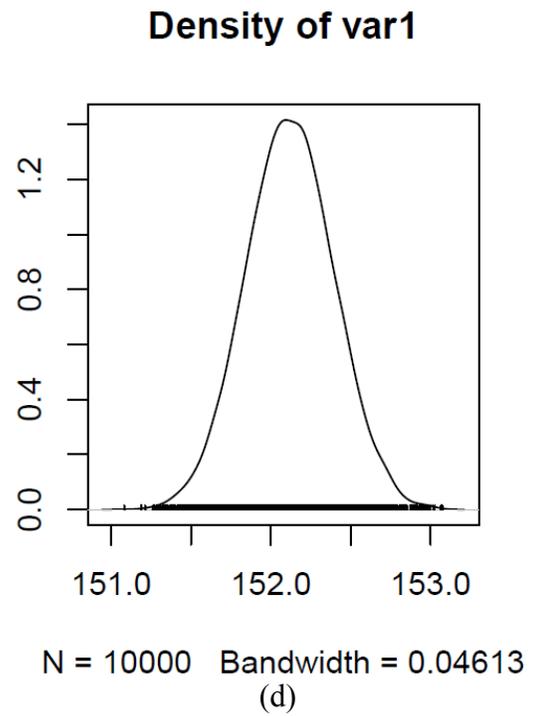
(d)

Figure 2

# Tables

**Table 1.** Summary of available list arguments for the function argument `data.values` of the function `bayes.regress()`

| List Argument | Description |
| --- | --- |
| `xtx` | The square matrix $X^T X$ of dimension $(k+1) \times (k+1)$, where $k$ = number of predictors; based on the summaries of the subset data, as shown in Formula (5) above. Default = 1. |
| `xty` | The square matrix $X^T Y$ of dimension $(k+1) \times (k+1)$, where $k$ = number of predictors; based on the summaries of the subset data, as shown in Formula (6) above. Default = 1. |
| `yty` | The square matrix $Y^T Y$ of dimension $(k+1) \times (k+1)$, where $k$ = number of predictors; based on the summaries of the subset data, as shown in Formula (7) above. Default = 1. |
| `xtx.inv` | Optional argument. This will be used for the Uniform prior distribution, to speed computations. If not provided, the argument `xtx` will be used to compute the inverse. |
| `numsamp.data` | The total number of data points in the full data set. Default = 1,000. |

**Table 2.** Summary of available list arguments for the function argument `beta.prior` of the function `bayes.regress()`. Note that all matrices must be symmetric and positive definite.

| Beta prior: Uniform | |
|---|---|
| **List Argument** | **Description** |
| `type ="flat"` | Indicates the Uniform prior. |

| Beta prior: Multivariate Normal prior with known mean vector $\mu$ and known covariance matrix C | |
|---|---|
| **List Argument** | **Description** |
| `type="mvnorm.known"` | Indicates the Multivariate Normal prior with known mean and known covariance. |
| `mean.mu` | The fixed known prior mean vector $\mu$ for the Multivariate Normal prior of $\beta$. The default is a vector of 0's with length equal to the length of $\beta$. |
| `cov.C` | The fixed known prior covariance matrix $C$ for the Multivariate Normal prior of $\beta$. The default is an identity matrix with dimension equal to the length of $\beta$. |
| `prec.Cinv` | The fixed known prior precision matrix for the Multivariate Normal prior of $\beta$. This is automatically calculated by R as the inverse of $C$ and does not need to be specified by the user. If provided, the preference will be given to this matrix. |

**Table 3.** Summary of available list arguments for the function argument `beta.prior` of the function `bayes.regress()`. Note that all matrices must be symmetric and positive definite.

| **Beta prior: Multivariate Normal prior with unknown mean vector $\mu$ and unknown covariance matrix $C$** ||
|---|---|
| **List Argument** | **Description** |
| `type="mvnorm.unknown"` | Indicates the Multivariate Normal prior with unknown mean and unknown covariance matrix. |
| `mu.hyper.mean.eta` | The fixed known hyperparameter mean vector $\eta$ for the Multivariate Normal hyperprior mean $\mu$. The default is a vector of 0's with length equal to the length of $\beta$. |
| `mu.hyper.prec.Dinv` | The fixed known hyperparameter precision matrix $D^{-1}$ for the Multivariate Normal hyperprior mean $\mu$. The default is an identity matrix with dimension equal to the length of $\beta$. |
| `Cinv.hyper.df.lambda` | The fixed known degrees of freedom $\lambda$ for the Wishart hyperprior for $C^{-1}$. The default value is the length of $\beta$. |
| `Cinv.hyper.invscale.Vinv` | The fixed known hyperparameter inverse scale matrix $V^{-1}$ for the Wishart hyperprior for $C^{-1}$. The default is an identity matrix with dimension equal to the length of $\beta$. |
| `mu.init` | The initial value for $\mu$ for the MCMC chain. The default is a vector of 1's with length equal to the length of $\beta$. |
| `Cinv.init` | The initial value for $C^{-1}$ for the MCMC chain. The default is an identity matrix with dimension equal to the length of $\beta$. |

**Table 4.** Summary of available list arguments for the function argument `sigmasq.prior` of the function `bayes.regress()`.

| Sigma squared prior: Inverse Gamma prior | |
|---|---|
| **List Argument** | **Description** |
| `type = "inverse.gamma"` | Indicates the Inverse Gamma prior. |
| `inverse.gamma.a` | The shape parameter *a* for the Inverse Gamma prior distribution, assumed known. Default = 1. |
| `inverse.gamma.b` | The scale parameter *b* for the Inverse Gamma prior distribution, assumed known. Default = 1. |
| `sigmasq.init` | The initial value for $\sigma^2$ parameter for the MCMC chain. Default = 1. |

| Sigma squared prior: Inverse Sigma Squared prior | |
|---|---|
| **List Argument** | **Description** |
| `type = "sigmasq.inverse"` | Indicates the $\left(\sigma^2\right)^{-1}$ prior. |
| `sigmasq.init` | The initial value for the unknown $\sigma^2$ parameter for the MCMC chain. Default = 1. |

**Table 5.** Posterior mean and posterior 2.5%, 50%, 97.5% percentiles for the unknown model parameters for the simulation and real airlines data sets.

| Linear Regression Data Set | Parameter | True Value (Simulated) | Posterior Mean | Posterior 2.5% | Posterior 50% | Posterior 97.5% |
|---|---|---|---|---|---|---|
| **Simulation Data** | $\beta_0$ | 0.4623 | 0.4624 | 0.4622 | 0.4624 | 0.4626 |
| | $\beta_1$ | -0.8638 | -0.8638 | -0.8640 | -0.8638 | -0.8636 |
| | $\beta_2$ | 1.4790 | 1.4791 | 1.4790 | 1.4791 | 1.4793 |
| | $\beta_3$ | -0.5139 | -0.5138 | -0.5140 | -0.5138 | -0.5136 |
| | $\beta_4$ | 0.4335 | 0.4337 | 0.4335 | 0.4337 | 0.4339 |
| | $\beta_5$ | 1.7971 | 1.7972 | 1.7970 | 1.7972 | 1.7973 |
| | $\beta_6$ | -0.0874 | -0.0875 | -0.0877 | -0.0875 | -0.0873 |
| | $\beta_7$ | -0.3138 | -0.3138 | -0.3140 | -0.3138 | -0.3136 |
| | $\beta_8$ | -0.8459 | -0.8457 | -0.8459 | -0.8457 | -0.8455 |
| | $\beta_9$ | 1.4213 | 1.4211 | 1.4210 | 1.4211 | 1.4213 |
| | $\beta_{10}$ | 1.6152 | 1.6153 | 1.6151 | 1.6153 | 1.6155 |
| | $\sigma^2$ | 1.0 | 1.0001 | 0.9999 | 1.0001 | 1.0004 |
| **Real Airlines Data** | $\beta_0$ | NA | 6.8482 | 6.7482 | 6.8481 | 6.9484 |
| | $\beta_1$ | NA | 0.8447 | 0.8435 | 0.8447 | 0.8458 |
| | $\beta_2$ | NA | -0.00088 | -0.00095 | -0.00088 | -0.00081 |
| | $\sigma^2$ | NA | 152.1 | 151.6 | 152.1 | 152.7 |

**Table 6.** Time (in seconds) to read in data and calculate summary statistics using `read.regress.data.ff()` function, and for producing 11,000 MCMC posterior samples (including burnin) using `bayes.regress()` function for specific choices of prior distributions. Note that $\beta$ prior 3) contains three sets of unknown model parameters, including an unknown precision matrix, which adds considerably to the computational time for producing MCMC posterior samples. The results are based on a computer with operating system Windows 7 64-bit and an Intel Core i7-4600U CPU 2.1 GHz Processor.

| No. of Pred-ictors | No. of Data Values | File Size | Time for Reading Data and Calculating Summary Statistics (seconds) | Time for Producing 11,000 MCMC Samples (seconds); $\beta$ prior 1), $\sigma^2$ prior 1) | Time for Producing 11,000 MCMC Samples (seconds); $\beta$ prior 2), $\sigma^2$ prior 2) | Time for Producing 11,000 MCMC Samples (seconds); $\beta$ prior 3), $\sigma^2$ prior 2) |
|---|---|---|---|---|---|---|
| 10 | $10^6$ | 104.3 MB | 8.47 | 0.45 | 0.75 | 1.86 |
| | $10^7$ | 1.043 GB | 90.26 | 0.47 | 0.76 | 1.82 |
| | $10^8$ | 10.43 GB | 1,331.02 | 0.48 | 0.73 | 1.84 |
| | $10^9$ | 104.3 GB | 16,495.18 | 0.48 | 0.76 | 1.86 |
| 100 | $10^5$ | 94.96 MB | 8.06 | 2.93 | 8.01 | 54.54 |
| | $10^6$ | 949.6 MB | 150.59 | 3.20 | 7.90 | 48.72 |
| | $10^7$ | 9.496 GB | 1,662.66 | 2.92 | 7.91 | 49.58 |
| | $10^8$ | 94.96 GB | 13,536.02 | 3.20 | 7.98 | 49.25 |
| 1000 | $10^4$ | 94.0 MB | 20.69 | 1,858.45 | 5.763.67 | 28,132.50 |
| | $10^5$ | 940 MB | 168.67 | 1,852.76 | 5,879.61 | 28,156.70 |
| | $10^6$ | 9.40 GB | 2,116.60 | 1,869.45 | 5,884.56 | 28,616.67 |
| | $10^7$ | 94.0 GB | 26,805.78 | 1,868.79 | 5,784.02 | 28,324.44 |